%Paper: hep-lat/9312073
%From: a71355a <a71355a@kyu-cc.cc.kyushu-u.ac.jp>
%Date: Sat, 18 Dec 93 12:07:27 JST

%nosupertex
%noinput
%%%%%%%%%%%%%%%%%%%%%%%%%%%%%%%%%%%%%%%%%%%%%%%%%%%%%%%%%%%%%%%%%%%%%
%
% This is the 1st version of reference macros
%                                              9 Aug 1987   K Nemoto
%         no use of file for keepref          11 Aug 1987
%         checker of redefinition of keepref  23 Aug 1987
%         The latest time stamp of the source "refmacro.tex"
%                                             87-08-24 16:41
%
%    Minimum Setup : can be used independently of "PHYZZX" (YRS)
%
%%%%%%%%%%%%%%%%%%%%%%%%%%%%%%%%%%%%%%%%%%%%%%%%%%%%%%%%%%%%%%%%%%%%%
\catcode `\@=11
\let\rel@x=\relax
\let\n@expand=\relax
\def\pr@tect{\let\n@expand=\noexpand}
\let\protect=\pr@tect
\let\gl@bal=\global
%%%%%%%%%%%%%%%%%%%%%%%%%%%%%%%%%%%%%%%%%%%%%%%%%%%%%%%%%%%%%%%%%
%
% First we define the macros required for sorting reference-numbers
%
%\input sort.tex
%
% Followings are the list macros explained in TeX book p378
%
\newtoks\t@a \newtoks\t@b \newtoks\next@a
\newcount\num@i \newcount\num@j \newcount\num@k
\newcount\num@l \newcount\num@m \newcount\num@n
\long\def\l@append#1\to#2{\t@a={\\{#1}}\t@b=\expandafter{#2}%
                         \edef#2{\the\t@a\the\t@b}}
\long\def\r@append#1\to#2{\t@a={\\{#1}}\t@b=\expandafter{#2}%
                         \edef#2{\the\t@b\the\t@a}}
\def\l@op#1\to#2{\expandafter\l@opoff#1\l@opoff#1#2}
\long\def\l@opoff\\#1#2\l@opoff#3#4{\def#4{#1}\def#3{#2}}
%
% "sort@@" macros sort the given number-list and make reference-number
%
\newif\ifnum@loop \newif\ifnum@first \newif\ifnum@last
\def\sort@@#1{\num@firsttrue\num@lasttrue\sort@t#1}
\def\sort@t#1{\pop@@#1\to\num@i\rel@x%\message{numi**\the\num@i}%
            \ifnum\num@i=0 \num@lastfalse\let\next@a\rel@x%
            \else\num@looptrue%
                 \loop\pop@@#1\to\num@j\rel@x%\message{numj=\the\num@j}%
                    \ifnum\num@j=0 \num@loopfalse%
                    \else\ifnum\num@i>\num@j%
                         \num@k=\num@j\num@j=\num@i\num@i=\num@k%
                         \fi%
                    \fi%
                    \push@\num@j\to#1%
                  \ifnum@loop\repeat%
                  \let\next@a\sort@t%
            \fi%
%            \message{find>>>>\the\num@i}%
            \print@num%
            \next@a#1}
\def\print@num{%
              \ifnum@first%
                 \num@firstfalse\num@n=\num@i\number\num@i%
              \else%
                 \num@m=\num@i\advance\num@m by-\num@l%
                 \ifcase\num@m\message{%
                   *** WARNING *** Reference number %
                   [\the\num@i] appears twice or more!}%
                 \or\rel@x%
                 \else\num@m=\num@l\advance\num@m by-\num@n%
                    \ifcase\num@m\rel@x%
                    \or,\number\num@l%
                    \else-\number\num@l%
                    \fi%
                    \ifnum@last\num@n=\num@i,\number\num@i\fi%
                 \fi%
              \fi%
              \num@l=\num@i%
              }
\def\pop@@#1\to#2{\l@op#1\to\z@@#2=\z@@}
\def\push@#1\to#2{\edef\z@@{\the#1}\expandafter\r@append\z@@\to#2}

%%%%%%%%%%%%%%%%%%%%%%%%%%%%%%%%%%%%%%%%%%%%%%%%%%%%%%%%%%%%%%%%%%%%%
%
% Next the main part of keepref macros
%
% The idea of \append@cs macro is described
%                 in the TeXBook EXERCISE 7.10 (p41)
% The idea of \if@first@use macro is described
%                 in the TeXBook EXERCISE 7.7  (p40)

\def\append@cs#1=#2#3{\xdef#1{\csname%
                    \expandafter\g@bble\string#2#3\endcsname}}
\def\g@bble#1{}

\def\if@first@use#1{\expandafter\ifx\csname\expandafter%
                              \g@bble\string#1text\endcsname\relax}
\def\keep@ref#1#2{\def#1{0}\append@cs\y@@=#1{text}\expandafter\edef\y@@{#2}}
\def\keepref#1#2{\if@first@use#1\keep@ref#1{#2}%
                 \else\message{%
                    \string#1 is redefined by \string\keepref! %
                    The result will be .... what can I say!!}%
                 \fi}

\def\Null{0}
\def\get@ref#1#2{\def#2{\string#1text}}
\def\findref@f#1{%
                \ifx#1\Null \get@ref#1\text@cc\R@F#1{{\text@cc}}%
                \else\rel@x\fi}

\def\findref#1{\findref@f#1\ref@mark{#1}}

\def\find@rs#1{\ifx#1\endrefs \let\next=\rel@x%
              \else\findref@f#1\r@append#1\to\void@@%
                    \let\next=\find@rs \fi \next}
\def\findrefs#1\endrefs{\def\void@@{}%
                    \find@rs#1\endrefs\r@append{0}\to\void@@%
                    \ref@mark{{\sort@@\void@@}}}
\let\endrefs=\rel@x

%%%%%%%%%%%%%%%%%%%%%%%%%%%%%%%%%%%%%%%%%%%%%%%%%%%%%%%%%%%%%%%%%%%%%
%
%   Hereafter a simplified version of reference macros in PHYZNEW.tex
%
\newcount\referencecount     \referencecount=0
\newif\ifreferenceopen       \newwrite\referencewrite
\newdimen\refindent          \refindent=30pt
%
%%%%%%%%%%%%%%%%%%PR%%%%%%%%%
\def\refmark#1{\attach{${}^{\scriptscriptstyle  #1)}$ }}
\def\NPrefmark#1{[#1]}
\def\ref@mark#1#2{\if#2,\rlap#2\refmark#1%
                  \else\if#2.\rlap#2\refmark#1%
                   \else\refmark{{#1}}#2\fi\fi}
\def\NPref@mark#1#2{\if#2,\NPrefmark#1,%
                  \else\if#2.\NPrefmark#1.%
                   \else\thinspace\NPrefmark{{#1}} #2\fi\fi}
\def\REF@NUM#1{\gl@bal\advance\referencecount by 1%
    \xdef#1{\the\referencecount}}
\def\R@F#1{\REF@NUM #1\R@F@WRITE}
\def\r@fitem#1{\par \hangafter=0 \hangindent=\refindent \Textindent{#1}}
%%%%%%%%%%%%%%%%%%%%%%%%%%%%%%%%%%%%
\def\itemref#1{\r@fitem{#1)}}
\def\NPrefitem#1{\r@fitem{[#1]}}
\def\NPrefs{\let\refmark=\NPrefmark \let\itemref=\NPrefitem%
            \let\ref@mark=\NPref@mark}
\def\PRrefs{\let\refmark=\attach}
\def\R@F@WRITE#1{\ifreferenceopen\else\gl@bal\referenceopentrue%
     \immediate\openout\referencewrite=\jobname.ref%
     \toks@={\begingroup \refoutspecials}%
     \immediate\write\referencewrite{\the\toks@}\fi%
    \immediate\write\referencewrite{\noexpand\itemref%
                                    {\the\referencecount}}%
    \immediate\write\referencewrite#1}
%

%%%%%%%%%  *************
\def\outrefs{%\par\penalty-400\vskip\chapterskip
   \spacecheck\referenceminspace
   \ifreferenceopen \toks0={\par\endgroup}%
   \immediate\write\referencewrite{\the\toks0}%
   \immediate\closeout\referencewrite%
   \referenceopenfalse \fi
%   \line{\fourteenrm\bf References\hfil}\vskip\headskip
   \centerline{\bf References}\vskip\headskip
   \input \jobname.ref
   }
\def\refoutspecials{\sfcode`\.=1000 \interlinepenalty=1000
         \rightskip=\z@ plus 1em minus \z@ }
%
% macros from "phyzzx"
%

%
\newskip\chapterskip	     \chapterskip=\bigskipamount
\newskip\sectionskip	     \sectionskip=\medskipamount
\newskip\headskip	         \headskip=8pt plus 3pt minus 3pt
\newdimen\chapterminspace    \chapterminspace=15pc
\newdimen\sectionminspace    \sectionminspace=10pc
\newdimen\referenceminspace  \referenceminspace=25pc
\def\Textindent#1{\noindent\llap{#1\enspace}\ignorespaces}
\def\space@ver#1{\let\@sf=\empty \ifmmode #1\else \ifhmode
   \edef\@sf{\spacefactor=\the\spacefactor}\unskip${}#1$\relax\fi\fi}
%%%%%%%%%%%%%%%%%%%
\def\attach#1{\space@ver{\strut^{\mkern 2mu #1)} }\@sf\ }
\def\spacecheck#1{\dimen@=\pagegoal\advance\dimen@ by -\pagetotal
   \ifdim\dimen@<#1 \ifdim\dimen@>0pt \vfil\break \fi\fi}
\catcode `\@=12
%%%%%%%%%%%%%%%hubref.tex%%%%%%
%%%%%%%%%%%%%%%%%%%%%%%%%%%%%%%%%%%%%%%%%%%%%%%%%%%%
%        Reference list
%       copied from TREF.TEX  Version Dec.2, 1989
%%%%%%%%%%%%%%%%%%%%%%%%%%%%%%%%%%%%%%%%%%%%%%%%%%%%
% Hubbard Model
%
\keepref\IH {M. Imada and Y. Hatsugai,
 J. Phys. Soc. Jpn. {\bf 58}(1989), 3752.}
\keepref\Sor {S.Sorella,E.Tosatti, S.Baroni,
 R.Car and M.Parrinello,
 Int. J. Mod. Phys.  {\bf B 1}(1988), 993.}
\keepref\Ya { C. N. Yang,
Phy. Rev. Letters  {\bf 63}(1989), 2144 .}
\keepref\Yb{ C. N. Yang,
 Rev. Mod. Phys.  {\bf 34}(1962), 694 .}
\keepref\WWZ {X.G. Wen, F. Wilczek and A. Zee,
 Phys. Rev. {\bf B39}(1989), 11413 .}
\keepref\BA {G.Baskaran and P.W.Anderson,
  Phys. Rev. {\bf B37}(1988), 580 .}
\keepref\NM{A. Nakamura and T. Matsui,
 Phys. Rev. {\bf B37}(1988), 7940 .}
\keepref\A { P.W. Anderson,
 Science{\bf 235}(1987), 1196.}
\keepref\AffM { I. Affleck and B. Marston,
 Phys. Rev.{\bf B37 }(1988), 3774.}
\keepref\Hub { J. Hubbard,
 Phys. Rev. Lett.{\bf 3}(1959), 77.}
\keepref\Hira { J.E. Hirsch,
 Phys. Rev.{\bf B28}(1983), 4059.}
\keepref\Hirb { J.E. Hirsch,
 J. of Stat. Phys.{\bf 43}(1986), 841.}
\keepref\Hirc { J.E. Hirsch,
 Phys. Rev.{\bf B31}(1985), 4403.}
\keepref\Hird { J.E. Hirsch,
 Phys. Rev. Lett.{\bf 51}(1983), 1900.}
\keepref\Paroa { A. Parola, S.Sorella, M.Parrinello and E.Tosatti,
 Phys. Rev.{\bf B43}(1991), 6190.}
\keepref\Parob { A. Parola, S.Sorella, S.Baroni,
M.Parrinello and E.Tosatti,
 Int. J. Mod. Phys.{\bf B3}(1989), 1865.}
\keepref\Trot { H.F. Trotter,
 Proc. Am. Math. Soc.{\bf 10}(1959), 545.}
\keepref\Suz { M. Suzuki,
 Prog. Theor. Phys.{\bf 56}(1976), 1454.}
\keepref\Loh { E.Y. Loh Jr., F.E.Gubernatis,
R.T.Scalettar, S.R.White,
D.J.Scalapino and R.L.Sugar ,
 Phys. Rev.{\bf B41}(1990), 9301.}
\keepref\FIa { N. Furukawa and M. Imada,
 J. Phys. Soc. Jpn{\bf 60}(1991), 810.}
\keepref\FIb { N. Furukawa and M. Imada,
 J. Phys. Soc. Jpn{\bf 60}(1991), 3669.}
\keepref\Kot {G. Kotliar ,
 Phys. Rev. {\bf B37}(1988), 3664.}
\keepref\Dag {E. Dagotto, A.Moreo, F.Ortolani, D.Poilblanc, J.Riera
and D.Scalapino,
 Phys. Rev. {\bf B45}(1992), 10741.}
\keepref\HatS {N. Hatano and M. Suzuki ,
 J. Phys. Soc. Jpn.{\bf 62}(1993), 847.}
\keepref\Tom {E.T. Tomboulis,
 Phys. Rev. Lett.{\bf 68}(1992), 3100.}
%%%%%%%%%%%%%%%%%%%% Chern-Simons %%%%%%%%%%%%
\keepref\Red { A.N. Redlich,
 Rhys. Rev. Lett.{\bf 52}(1984), 18.; Rhys. Rev. {\bf D29}(1984), 2366.}
%\keepref\Red { A.N. Redlich,
%Rhys. Rev. {\bf D29}(1984), 2366. }
\keepref\Ish { K. Ishikawa,
 Phys. Rev. Lett.{\bf 53}(1984), 1615.}
\keepref\So { H. So,
 Prog. Theor. Phys.{\bf 73}(1985), 528.; {\bf 74}(1985), 585.}
%\keepref\So { H. So,
% Prog. Theor. Phys.{\bf 74}(1985), 585.}
\keepref\FS { E. Fradkin and F.A. Shaposnik,
 Phys. Rev. Lett.{\bf 66}(1991), 276.}
%%%%%%%%%%%%%%%%%%%%%%%%%%%
%\keepref\ { ,
% {\bf }(19), .}
%

\PRrefs
%\input phyzzx
%\magnification=\magstephalf
%\magnification=1095
%\magnification=1313
\magnification=1200
%\scrollmode
%\nopagenumbers
\def \ref(#1){$^{#1)}$}
%%% def   \box
\def\sqr#1#2{{\vcenter{\vbox{\hrule height.#2pt \hbox{\vrule width.#2pt
height#1pt \kern#1pt \vrule width.#2pt}\hrule height.#2pt}}}}

%%%%Roman%%%%%%%%%%%%%

%\def \Romannumeral(#1) {\uppercase\expandafter{\romannumeral#1}}
\def \Rn#1 {\uppercase\expandafter{\romannumeral#1}}
%%%%%%%%%%%%%%%%%%%%%%%%%%%%
%%%%%%%%%%%%%%%%%%%%%
\def\gtsim{\mathrel{\hbox{\raise0.2ex
\hbox{$>$}\kern-0.75em\raise-0.9ex\hbox{$\sim$}}}}
\def\ltsim{\mathrel{\hbox{\raise0.2ex
\hbox{$<$}\kern-0.75em\raise-0.9ex\hbox{$\sim$}}}}
%%%%%%%
\let \q=`
\hsize=15.0 true cm
\vsize=22.0 true cm
\hoffset=0.5 true cm
\voffset=-0.5 true cm
%%%%%%%%%%%%%%%%%% \font\bigfa=cmr10 % scaled 1095
%%%%%%%%%%%% \font\bigfb=cmr10 % scaled 1313
%print
%\font\bigfc=cmbx12  % scaled 1574
%\font\bigfc=cmbx10 scaled  1315  %pc98preview
\font\bigfc=cmbx10 scaled  1200  %pc98preview
%\font\bigfc=cmbx10 scaled  1095   %pc-pr201
%%%%%%%%%%%%%%%\font\bigfc=cmbx10 scaled  1578 %pc98preview
%\baselineskip=15pt
\baselineskip=24pt plus 2pt minus 2 pt
\def \ref(#1){$^{#1)}$}                                %%%%%%%%%references
\def\Fig(#1){$${\overline{\underline{\rm \ Fig.{#1}\ }}}$$}  %%%%%%%Figures

\def\b{\beta}

\def\D{\Delta}
\def\d{\delta}
\def\e{\eta}
\def\ve{\varepsilon}

\def\M{M_f}

\def\r{\rho}
\def\s{\sigma}
\def \t{\tau}

\def \vc{\vert 0 \rangle}
\def \ket{\langle 0 \vert}
\def \tvc{\vert \tilde 0 \rangle}
\def \tket{\langle \tilde 0 \vert}
\def \ph{\phi}

\def \Romannumeral(#1) {\uppercase\expandafter{\romannumeral#1}}
\def \Rn(#1) {\uppercase\expandafter{\romannumeral#1}}

%\topskip 2cm

%
%\rightline{\hfill Version 3.2} \par
\rightline{\hfill KYUSHU-HET-11} \par
\rightline{\hfill SAGA-HE-49}\par
\hfill Nov., 1993\par
\vskip 3 true cm
%\bigskip
%%%%%%%%%%%%%%%%%%%%%%%%%%%%%%%%%%%%%%%%%%%%%%%%%%%%%%%%%%%%%%%%%%%%%%%%%
%title
  \centerline {\bigfc  Complex wave function,  Chiral spin order parameter}
  \centerline {\bigfc and}
%\centerline {\bigfc ``Phase Problem"  }
%ch
\centerline {\bigfc Phase Problem  }
\vskip 1 true cm
%author
\centerline {Masahiro IMACHI\footnote{*}{
e-mail:a71355a@kyu-cc.cc.kyushu-u.ac.jp}}
\smallskip
\centerline{and}
\smallskip
\centerline {Hiroshi YONEYAMA$^{\dagger)}$\footnote{**}{
e-mail:yoneyama@math.ms.saga-u.ac.jp}}
\bigskip
\centerline{\sl  Department of Physics, Kyushu University, Fukuoka 812,
Japan}
\centerline{\sl and}
\centerline{\sl  Department of Physics,
            Saga University, Saga 840, Japan$^{\dagger) }$ }
\vfill\eject
%% FOLLOWING LINE CANNOT BE BROKEN BEFORE 78 CHAR
%%abs%%%%%%%%%%%%%%%%%%%%%%%%%%%%%%%%%%%%%%%%%%%%%%%%%%%%%%%%%%%%%%%%%%%%%%%%%%
\vskip  1 true cm
\leftline\bf{ABSTRACT}
\smallskip
{\leftskip 30pt \rightskip 30pt
We study  the two dimensional Hubbard model by use of the ground state
algorithm in the  Monte Carlo simulation.
 We  employ  complex wave functions as trial
 function in order to have a close look at  properties such as chiral spin
order
($\chi$SO) and flux phase.  For  half filling, a particle-hole transformation
leads to  sum rules with respect to the Green's functions for a certain
choice of
 a set of wave functions.   It is then analytically
shown that the sum rules lead to the absence of the $\chi$SO.
 Upon doping, we are confronted with the sign problem, which in  our case
%ch
 appears as a  ``phase problem" due to the phase of the
Monte Carlo weights.
The average of the phase shows an
exponential decay as a function of inverse temperature similarly  to that of
sign by
Loh Jr. et. al. .  We compare the numerical results  with    those of exact
numerical calculations.

\par}
\vskip  1.5 true cm
\vfill\eject
%%%sec1%%%%%%%%%%%%%%%%%%%%%%%%%%%%%%%%%%%%%%%%%%%%%%%%%%%%%%%%%%%%%%%%%%%%
\leftline{\bf \S 1. Introduction } \par
\bigskip
  The two dimensional Hubbard model has recently attracted much attention
 in view of  possibly  providing  insight into the high-$T_c$ Cu-O
superconductors.
In the strong  $U$ (Coulomb repulsive force) limit at  half filling, the
model is
 equivalent to the antiferromagnetic
Heisenberg model, and  in good agreement with  the antiferromagnetic
behavior of superconducting materials.
When the system is doped, however, the ground state properties appear
less clear.
  Concerning  ground states of the strongly correlated electronic
 systems, various hypothetical states such as spin-liquid \findref \WWZ,
 resonating valence bond( RVB) \findrefs \A \BA \endrefs, and flux
phase\findref \AffM, etc.
are proposed.
Of  particular interest  is a possibility of the violation  of  P (parity)
and T (time)
 invariance  in such systems\findref \WWZ.  From the field theoretical point
of view also,
 the relevance of spontaneous breakdown  of such symmetries have been pointed
out, and it is conjectured  that   a Chern-Simons term in 2+1 dimensional
QED,\findrefs \Red \So \Ish \endrefs
 which is regarded as an effective theory describing the RVB picture,  may
violate the confinement of the fundamental charge ( the charge-spin
separation)\findref \FS.
 In spite of such  intriguing arguments, however, any  clear evidence  has
not so
 far been found. \par
 The above ideas  are  proposed   based  basically upon the mean
field approach,
and therefore the stability of its vacuum against  full quantum fluctuations
 is not clear at all.
In such situations numerical approach is a  suitable  method to study
the strongly correlated electron systems.
Among various approaches,  Monte Carlo simulation is one of the most
promising
methods.
 \par
  In the present paper we present a Monte Carlo study\findrefs
\Hira \IH  \endrefs of the two dimensional
Hubbard model.\findref \Hub  So far similar works have been done.  Although
our approach is
basically the same, and the size of lattice is also quite limited,
 our motivation of the study  is two fold.
  One is to use complex wave functions as a trial function
in the ground state algorithm in order to have a close look at the
properties stated above, namely, ones of the chiral spin order ($\chi$SO)
and flux
phase,\findrefs \AffM \Kot \endrefs where the complexity of the expectation
values
 is involved.
At half-filling, a particle-hole symmetry makes the theory   transparent
to look at.   By use of such symmetry, for some choice of a set of wave
functions
  (we call it ``dual choice''),  it is shown that there hold sum rules with
respect to
the Green's function with spin up and
that with down.   They are of different form  for  even and odd lattice
spacings.
 We analytically show that these sum rules lead to the absence  of the
expectation
value of the $\chi$SO parameter at half-filling\findref \Tom. \par
  Another motivation is to study an effect of such wave functions to the so
called
sign problem, which frequently arises  in the quantum Monte Carlo
simulations.
A widely used algorithm employs the Hubbard-Stratonovich
transformation\findrefs \Hira \IH \endrefs,
 rewriting  the system to the $2+1$ dimensional classical system of the Ising
 dynamical variable.  The conventional importance sampling with
configurations
of this variable encounters the negative weight which is originated from the
fermionic determinant.  This prevents  probabilistic  algorithm from
being applied to the system.  A way to circumvent the problem is to
use the positive weight and to estimate  expectation value of operator
combined
with the average value of the sign.   This however causes large
statistical  errors
 for some cases.\par
   The origin of the sign problem in the  framework of trial function
approach
 is not yet clear, and thus a way to overcome it  has not yet
 been established.
 Some improvements have been tried by employing optimized wave
function.\findref \FIb
The results depend very much on the choice of the wave functions of the trial
state. It is therefore not trivial at all what to obtain  if we employ
the complex
 wave function. In our case the sign of the weights are replaced
%ch
by  their phase (``phase problem") .
 We paid  a special attention to the average  of the phase.
In the present paper,   we study, as  doped cases, 14 fermion and 10 fermion
systems on 4$\times$4 lattice.  The latter case is almost free from
%ch
the phase  problem, but the former receives rather serious effect.
A measurement of the  $\chi$SO for the former case unfortunately gets so
large
 noises  to extrapolate to the low temperature limit.  However other physical
quantities such as one particle density operator and spin-spin correlations
are rather in good agreement with the result obtained by exact numerical
 calculations\findrefs \Paroa \Parob \Dag \endrefs.\par
    This paper is organized as follows. In the following section,
we present the
 formalism.
In section 3,  the half filled case is analytically discussed based upon
a particle-hole transformation. Some choice of a set of wave functions is
shown
to lead to a sum rule.   The  absence of the expectation value of $\chi$SO is
derived  at half filling.
In section 4, results of numerical simulations are presented for
half filled and
doped cases. Our conclusions and discussion are presented in section 5 .
\vfill\eject
%%%%%%%%%%%%%%%%%%%%%%%%%%%%%%%%%%%%%%%sec2%%%%%%%%%%%%%%%%%%%%%%
\leftline {\bf \S 2 Hubbard model and Monte Carlo algorithm}\par
   In this section, we shortly present the formalism for fixing
 notations as well as
 for making the paper self-contained. It is basically based on that of
Hirsch\findrefs \Hira \Hirb \Hirc \Hird \endrefs and
Imada-Hatsugai\findref \IH.\par
  The Hubbard hamiltonian is defined as
$$H=-t\sum_{\langle i j \rangle \s}c_{i \s}^\dagger c_{j \s}+h.c.
+U\sum_i n_{i \uparrow}n_{i \downarrow} \eqno(2.1)$$
where $c_{i \s} (c_{i \s}^\dagger)$ is an annihilation (creation) operator of
an
electron with spin $\s$ ($\s=$ up or down) at site $i$, and
$ n_{i \s} $ is the number operator at site $i$ of  spin   $\s$.
 The first term is the hopping
term of electron over   nearest neighbor sites $\langle i j \rangle$,
and the second stands for the on-site  repulsive Coulomb potential with
the strength $U$($>0$).
An algorithm to perform Monte Carlo simulations  here is so called ground
state
 algorithm for fixed number of fermions.  Expectation value of
a physical observable $O$  is calculated by
$$\langle O \rangle={\langle\Phi\vert O\exp(-\b H)\vert\Phi\rangle
                              \over   \rho(\b;\Phi)} \eqno(2.2)$$
where
$$\rho(\b;\Phi)\equiv \langle\Phi\vert \exp(-\b H)\vert\Phi\rangle
\eqno(2.3)$$
is given in terms of an appropriate trial state vector $ \vert\Phi\rangle$.
As $\b$ goes to infinity, (2.2) gives an expectation value for true ground
state
$\vert \psi_0\rangle$,
unless the trial function is orthogonal to the true ground state;
$$\exp(-\b H)\vert \Phi \rangle \rightarrow
\exp(-\b E_0)\vert \psi_0 \rangle\langle\psi_0\vert\Phi\rangle \eqno(2.4)$$
For evaluating (2.2) we make Monte Carlo simulations.
Introducing discrete pseudotime with interval $\D\t$, $\rho(\b;\Phi)$ is
rewritten as
$$\rho(\b;\Phi)=\langle\Phi\vert (\exp(-\D\t H))^L\vert\Phi\rangle
\eqno(2.5)$$
where $\D \t=\b /L$ and $L$ denotes the Trotter size, i.e., the number  of
time slices.
  Separating the hopping term $H_0$ and the
self-interaction term $H_1$ of the hamiltonian,
one introduces Trotter-Suzuki formula\findrefs \Trot \Suz \endrefs,
%
%$$ \exp (-\D\t H)= \exp({-\D\t \over 2} H_0) \exp(-\D \t H_1)
%             \exp({-\D\t \over 2} H_0)+O((\D \t)^3) \eqno(2.6) $$

$$\exp(-\D\t H)=\exp({-\D\t \over 2} H_0)\exp(-\D \t H_1)
              \exp({-\D\t \over 2} H_0)+O((\D\t)^3) \eqno(2.6) $$
To deal with the quartic interaction term $H_1$ of the fermion field,
one introduces the Hubbard-Stratonovich (H-S) transformation of the Ising
type
for each site.
$$\exp(-c n_{\uparrow}n_{\downarrow})
={1 \over 2} \sum_{s=\pm 1}\exp\, (2\,a\,  s(n_{\uparrow}-n_{\downarrow})
  -{c \over 2}(n_{\uparrow}+n_{\downarrow})) \eqno(2.7)$$
where $a=\tanh^{-1}\sqrt{\tanh(c/4)}$ for a constant $c$.
Eqs.(2.5), (2.6) and (2.7) lead to
$$
\rho(\b;\Phi) = \sum_{\lbrace s_{i l}\rbrace} \langle \Phi \vert \,
\prod^L_{l=1}\,
                               (\, w_0 \, w_1(\lbrace s_{il} \rbrace)\, w_0
\, ) \, \vert \Phi \rangle
 \eqno(2.8)$$
where the summation is taken over the H-S variable $s_{i l}$ on the 2+1
 dimensional lattice.
Quantities $w_0$ and $ w_1(\lbrace s_{il} \rbrace) $ are defined as
$$\eqalign{
w_0  =&\exp(-\D\t H_0/2) \cr
   w_1(\lbrace s_{il} \rbrace) =& \prod^N_{i=1}\,({1 \over 2}\, \exp
\,\lbrack \,2 \,a_U \,s_{il}
\,(n_{\uparrow}-n_{\downarrow})-{\D \t U \over 2}
(n_{\uparrow}+n_{\downarrow})\rbrack ) \cr
} \eqno(2.9)
$$
where $a_U=\tanh^{-1}\sqrt{\tanh(\D \t U/4)}$, and $N$ is the number of
lattice
sites.\par
  We assume a factorization of spin up and down sectors  for the trial state,
$$
\vert\Phi\rangle = \vert \Phi_\uparrow\rangle
\vert \Phi_\downarrow \rangle \eqno(2.10)$$
and   take   wave functions of single particle basis for $\M$ fermions for
each
spin sector
$$
\vert \Phi_\s \rangle
=\prod_m^{\M }\, (\, \sum_i^N \, (\phi_{\s})_{m i}\, c_{i \s}^\dagger)\,\vert
0 \rangle
\eqno(2.11)
$$
where $(\phi_{\s})_{m i} (i=1,...,N; m=1,...,\M)$ is a  $\M \times N$
matrix.  In this paper we allow $(\phi_{\s})_{m i}$ to be complex in general.
Using   (2.11),    $\rho(\b;\Phi)$ (2.8),  is explicitly represented in a
matrix form
$$
 \eqalign{\rho(\b;\Phi)=& \sum_{\lbrace s_{i l} \rbrace} W(\b;s) \cr
                 W(\b;s)=&W_\uparrow(\b;s)W_\downarrow(\b;s)}\eqno(2.12)
$$
where  $W_\s(\b;s)$ is given by a determinant of a product of matrices
$$
W_\s(\b;s)={\rm det} \, \lbrack \phi_\s^\dagger \, \prod_l^L (M_0 \,
M_{1\s}(l) \, M_0) \,
\phi_\s \, \rbrack\eqno(2.13)
$$
Matrix  $M_0$ is given by
$$\eqalign{
M_0=&\exp (-K) \cr
K_{i j}=& \cases{ -\D \t t/2      & {\rm nearest neighbor} (i,j) \cr
                                  0                   & {\rm otherwise} \cr}
\cr
}\eqno(2.14)$$
while  $M_1$, which depends upon  the H-S variable,
 is a diagonal matrix with  element
$$ (M_{1\s}(l))_{i i}=\exp \, [\, \s \, a_U \, s_{i l} \, - \D \t /2 \,]
\eqno(2.15)$$
for $\s=+1 (-1)$  for spin up (down).\par
   According to Imada-Hatsugai,\findref \IH single particle
Green's function  at the time slice $l$
$$(\, G_\s (l) \, )_{i j}\equiv {\langle L_\s (l) \vert \, c_{j \s}^\dagger\,
c_{i \s}
                             \, \vert R_\s(l) \rangle \over \langle L_\s (l)
\vert R_\s(l) \rangle}
\eqno(2.16)$$
is expressed as
$$
G_\s (l)=R_\s (l) g_\s (l) L_\s ^\dagger(l)\eqno(2.17)
$$
where
$$
g_\s (l)=( L_\s^\dagger (l) R_\s (l))^{-1} \eqno(2.18)
$$
Matrix $R_\s (l)$ is $N \times \M$ matrix, and $L_\s^\dagger (l)$ is $\M
\times N$
 one;
$$\eqalign{
R_\s (l)=&M_{1\s}(l)M_0\prod_{l=l+1}^L (M_0 M_{1\s}(l) M_0) \phi_\s \cr
 L_\s ^\dagger(l)=&\phi_\s^\dagger \,\prod_{l=1}^{l-1} \,(\,M_0 M_{1\s}(l)
M_0\,) \, M_0 \cr
}\eqno(2.19)$$
\par
   Finally,  expectation value of operator  with spin $\s$ for the ground
state  is
 averaged at each time slice, and given by
$$
\eqalign{
\langle O_\s \rangle
  =&{ {1 \over L} \sum_l \sum_{\lbrace s \rbrace } {\rm det}\,(\,L_\s
(l)^\dagger\,
                O_\s\, R_\s (l)\,) \,W_{-\s} \over  \sum_{\lbrace s\rbrace} W
} \cr
 =&{1 \over L} \sum_{  l \lbrace s \rbrace}{\, {\rm det}\, (L_\s
(l)^\dagger\,
O_\s\, R_\s (l)\,) \over  W_\s}  W
{1 \over \sum_{\lbrace s\rbrace} W} \cr }\eqno(2.20)
$$
  We employ the heat-bath method for updating the H-S variables.
 $W$, however,  may not be real  in some cases, though $\rho(\b;\Phi)$
is positive semi-definite.   These  cases lead one to so called sign problem.
In our case it is replaced by that of the  phase, which prevents us from
doing
importance sampling.
Conventional way out of it is to define a positive measure
$$
 P_+ ={\vert W \vert \over \sum_{{s}} \vert W \vert} \eqno(2.21)
$$
and evaluate (2.20) in such a way as
$$\eqalign{
\langle O_\s \rangle
=&({1 \over L} \sum_{  l \lbrace s \rbrace}{{\rm det}(L_\s (l)^\dagger
O_\s R_\s (l)) \over  W_\s}  \vert W \vert e^{i \theta_w})
{1 \over \sum_{\lbrace s\rbrace} \vert W \vert e^{i \theta_w}} \cr
  =&({1 \over L} \sum_{  l \lbrace s \rbrace}{{\rm det}(L_\s (l)^\dagger
O_\s R_\s (l)) \over  W_\s}  \vert W \vert e^{i \theta_w}
/\sum_{\lbrace s\rbrace} \vert W \vert)
({1 \over \sum_{\lbrace s\rbrace} \vert W \vert e^{i \theta_w}
 /\sum_{\lbrace s\rbrace} \vert W \vert})
\cr
=& \langle {1 \over L} \sum_{  l }{{\rm det}(L_\s (l)^\dagger
O_\s R_\s (l)) \over  W_\s}  e^{i \theta_w} \rangle_+
/
  \langle e^{i \theta_w}\rangle_+ \cr
}\eqno(2.22)$$
where $W \equiv \vert W \vert e^{i \theta_w}$, and
$ \langle \bullet  \rangle_+ $ is the expectation value of $\bullet$ with the
positive measure $P_+$.

\bigskip
%%%%%%%%%%%sec3%%%%%%%%%%%%%%%%%%%%%%%%%%%%%%%%%%%%%%%%
\leftline{\bf \S 3 Dual choice of wave function and sum rule}\par
 The hamiltonian of the system respects a symmetry under particle-hole
 transformation
$$ c_{i \s}^\dagger \rightarrow   \eta^i d_{i -\s}, \eqno(3.1)$$
where $d_{i -\s}$ denotes the annihilation operator of a hole at site $i$
with spin
$-\s$ .  The sign factor $\eta^i=(-1)^i$ is positive (negative) for even
(odd) sites.
In this section we will show the Green's functions satisfy sum rules when the
system
 is half-filled, and ``dual choice" of the wave function is made. The sum
rule is satisfied
when the particle-hole symmetry holds.
The sum rules are
$$ f_{i j}= \cases{
                    \d_{i j}-g_{j i}^*    &  for $|i-j|=$even \cr
                     g_{j i}^*                 &       for $|i-j|=$odd \cr }
\eqno(3.2)$$
where $f_{i j}$ and $g_{i j}$ denote the spin up and down Green's function in
the
configuration space, respectively,  and $*$ is the complex conjugation.
\smallskip
%%%%%%%%%%%sec31%%%%%%%%%%%%%%%%%%%%%%%%%%%%%%%%%%%%%%%%
\leftline{\it 3-1 particle-hole transformation}\par
Particle-hole (p-h) transformation gives the correspondence
$$     \vc                  \longleftrightarrow              |\tilde 2\rangle
\qquad
         |1\s\rangle       \longleftrightarrow        |\tilde 1 \s\rangle
\qquad
         |2\rangle           \longleftrightarrow        |\tilde 0\rangle$$
at each site, where $\vc, |1\s\rangle$ and $|2\rangle $denotes zero, single
particle
with spin $\s$ and two particles (spin up and down),
while   $ |\tilde 0\rangle, |\tilde 1 \s\rangle$ and
  $|\tilde 2\rangle$denotes
 zero hole, single hole with spin $\s$ and two holes, respectively.
  Note that hole states are denoted with tilde.

The creation operator of a particle with quantum number $m$ (e.g.,momentum)
 is given as
$$ c_{m \s}^\dagger = \sum_j \ph_{m j} c_{j \s}^\dagger\eqno(3.3)$$
where  $\ph_{m j}$ is  the wave function.  Suffices $m$ and $j$ are for
example
abbreviated  as
$m=k=(k_1,k_2),  j=n=(n_1,n_2)$
for the momentum $(k_1,k_2)$ and the position $(n_1,n_2)$.
Let us see how states transform under the p-h transformation.
For illustration we consider one-dimensional half-filled ($\M=2$) system with
$N=4$ (4 site system).
Spin $\s$ state is
 $$ c_{m_1 \s}^\dagger c_{m_2 \s}^\dagger \vc
        =\sum_{j_1 j_2} \ph_{m_1 j_1} \ph_{m_2 j_2}
 c_{j_1 \s}^\dagger c_{j_2 \s}^\dagger \vc\eqno(3.4)$$
where $m_1$ and $m_2$ are quantum numbers of the two particles with spin
$\s$.
Under the p-h transformation, this state goes to
$$ \sum_{j_1 j_2} \ph_{m_1 j_1} \ph_{m_2 j_2}\e_{j_1}\e_{j_2}
d_{j_1 -\s} d_{j_2 -\s}
d_{1 -\s}^\dagger d_{2 -\s}^\dagger d_{3 -\s}^\dagger d_{4 -\s}^\dagger
d_{1 \s}^\dagger d_{2 \s}^\dagger d_{3 \s}^\dagger d_{4 \s}^\dagger
\tvc\eqno(3.5)$$
We pay an attention to spin $-\s$ quantities, and then rewrite
$$d_{1 -\s}^\dagger d_{2 -\s}^\dagger d_{3 -\s}^\dagger d_{4 -\s}^\dagger
={1 \over 4!} \sum_{l_1 l_2 l_3 l_4} \ve_ {l_1 l_2 l_3 l_4}
d_{l_1 -\s}^\dagger d_{l_2 -\s}^\dagger d_{l_3 -\s}^\dagger d_{l_4
-\s}^\dagger\eqno(3.6)$$
where $l_1,..,l_4$ denote the sites and $\ve_ {l_1 l_2 l_3 l_4}$ stands for
the Levi-Civita symbol.
The quantities with $-\s$ in (3.5) is
$$\eqalign{
\sum_{j_1 j_2 l_1 l_2 l_3 l_4}& {6 \over 4!}\ph_{m_1 j_1} \ph_{m_2
j_2}\e_{j_1}\e_{j_2}
\ve_ {l_1 l_2 l_3 l_4}(\d_{j_2 l_1}\d_{j_1 l_2}-\d_{j_2 l_2}\d_{j_1 l_1})
d_{l_3 -\s}^\dagger d_{l_4 -\s}^\dagger\tvc \cr
=&-{12 \over 4!} \sum_{l_1 l_2 l_3 l_4} \ph_{m_1l_1} \ph_{m_2
l_2}\e_{l_1}\e_{l_2}
\ve_ {l_1 l_2 l_3 l_4}d_{l_3 -\s}^\dagger d_{l_4 -\s}^\dagger\tvc \cr
=&-{1 \over 2}\D \sum_{j_3 j_4 m_3 m_4}\ph_{m_3 j_3}^\ast \ph_{m_4 j_4}^\ast
\e_{j_3}\e_{j_4}\ve_ {m_1 m_2 m_3 m_4}d_{j_3 -\s}^\dagger d_{j_4
-\s}^\dagger\tvc}\eqno(3.7)$$
%%%%%%%ch eta_j3 is inserted in eq.3.7%%%%%%%%%%%%%%%%%%%%%%%
where $\D=$ det$\tilde \ph$ and $\tilde \ph_{m j}=\ph_{m j} \e_j$.
Here we assumed the orthonormal condition of  wave functions
$$\eqalign{
\sum_j \tilde \ph_{m j} \tilde \ph_{m' j}^\ast=&
\sum_j  \ph_{m j}  \ph_{m' j}^\ast \e_j^2 =\d_{m m'} \cr
\sum_m \tilde \ph_{m j} \tilde \ph_{m j'}^\ast=&
\sum_m  \ph_{m j}  \ph_{m j'}^\ast \e_j \e_j'=\d_{j j'}\cr
}\eqno(3.8)$$
and used an identity
$$\ve_ {j_1 j_2 j_3 j_4} \tilde \ph_{m_1 j_1} \tilde\ph_{m_2 j_2}
\tilde \ph_{m_3 j_3} \tilde\ph_{m_4 j_4} =\ve_ {m_1 m_2 m_3 m_4} \D
\eqno(3.9)$$
{}From Eq.(3.8) and (3.9)  it follows that
$$\ve_ {j_1 j_2 j_3 j_4} \tilde \ph_{m_3 j_3} \tilde\ph_{m_4 j_4} =
\tilde \ph_{m_1 j_1}^\ast \tilde\ph_{m_2 j_2}^\ast \ve_ {m_1 m_2 m_3 m_4} \D
\eqno(3.10)$$
which leads to (3.7).
Note that the quantum numbers $m_3, m_4$ on the r.h.s. of (3.7) turn out to
be
different  from the ones (fixed) $m_1, m_2$ on the l.h.s. of (3.5).  It is
due to the
 factor $\ve_ {m_1 m_2 m_3 m_4}$, or the Pauli's principle.
The consequence of (3.9) is ;
this state is similar to the original one  (l.h.s. of (3.4)) if we replace
(i) $c^\dagger$ by  $d^\dagger$, (ii) $\s$ by $-\s$, (iii) $\ph_{m j}$ by
$\tilde\ph_{\bar m j}^\ast=\ph_{\bar m j}^\ast \e_j $ ,where $\bar m$ is the
coset of $m$, namely a combination which is completely different from the
original set of quantum number $m$.
Having in mind the fact that the hamiltonian is symmetric under p-h
transformation, the original Green's function of spin $\s$ based on the wave
function $\ph_{m j}$ is transformed to that of spin $-\s$ based on
 $\tilde \ph_{\bar m j}^\ast$.
\smallskip
%%sec32%%%%%%%%%%%%%%%%%
\leftline{\it 3-2 dual choice}\par
Consider a set of wave function $\lbrace \ph_{m j} \rbrace$ which defines
(3.6)
of single particle
basis.   We learned  in the previous sub-section that the corresponding hole
 basis is  $\lbrace \ph_{\bar m j}^\ast \e_j \rbrace$. When the set
 $\lbrace \ph_{\bar m j} \e_j \rbrace $ coincides with the
%ch
 set
$ \lbrace \ph_{m j}\rbrace$, that is,
when the hole basis wave function in (3.9) is simply
complex conjugate of the particle basis wave function, we call the  choice of
the
wave functions $\lbrace \ph_{m j}\rbrace $ `` dual choice ''.  Let us take
the previous
 example of the  one dimensional 4-site case.
We can take  $\ph_{m j}$ as
$$\ph_{1 j}=1 \qquad \ph_{2 j}=C \exp (i 2\pi n/4)\eqno(3.11)$$
where $n$ is an integer standing for the location $j$, and $C$ is a
normalization
 constant. %ch
   Then $\ph_{\bar m j} $ can be taken as
$$\ph_{\bar 1 j}=C \exp (-i 2\pi n/4)      \qquad
\ph_{\bar 2 j}=C \exp (i 4\pi n/4).\eqno(3.12)$$
%ch
We see that
$$\ph_{\bar 1 j}^\ast \e_j =\ph_{2 j}^\ast    \qquad
    \ph_{\bar 2 j}^\ast \e_j =\ph_{1 j}^\ast,\eqno(3.13)$$
%ch
and it shows the choice (3.11) is a ``dual choice".
\smallskip
%%%%%sec33%%%%%%%%%%%%%%%%%%%%%%%%%%%%%%%%%
\leftline{\it 3-3 sum rule}\par
Since the hamiltonian is invariant under the p-h transformation, we drop  for
simplicity the Boltzmann factor $\exp(-\b H)$ in the following arguments.
%ch%%%%%here comes a rather long correction%%%%%%%%%%%
Due to the dual choice of wave function, matrix element of $c$'s is
transformed to that of $d$'s with the same set of wave functions except for
complex conjugation.
$$\eqalign{
\ket c_{m_2 \s} c_{m_1 \s} & c_{m_1 \s}^\dagger c_{m_2 \s}^\dagger \vc \cr
=&\sum_{j_1j_2 j'_1 j'_2} \tket d_{j'_2 -\s} d_{j'_1 -\s} \ph_{m_1
j'_1}\ph_{m_2 j'_2}
\ph_{m_1 j_1}^\ast \ph_{m_2 j_2}^\ast d_{j_1-\s}^\dagger d_{j_2 -\s}^\dagger
\tvc}\eqno(3.14)$$
 The Green's function is changed to
$$\eqalign{
G_{\s i j}({\ph})=&
\ket c_{m_2 \s} c_{m_1 \s} c_{i \s} c_{j \s}^\dagger c_{m_1 \s}^\dagger
 c_{m_2 \s}^\dagger \vc \cr
=& \tket d_{j'_2 -\s} d_{j'_1 -\s} \ph_{m_1 j'_1}\ph_{m_2 j'_2}
\e_i d_{i -\s}^\dagger \e_j d_{j -\s}
\ph_{m_1 j_1}^\ast \ph_{m_2 j_2}^\ast d_{j_1-\s}^\dagger d_{j_2 -\s}^\dagger
\tvc \cr
=&\d_{i j} - \e_i \e_j G_{-\s j i}({\ph^\ast}) \cr
=&\d_{i j} - \e_i \e_j G_{-\s j i}^\ast ({\ph}) \cr
}\eqno(3.15)$$
where the reality of $G$ is used.
We therefore have
$$f_{i j}= \cases {
                \d_{i j}-g_{j i}^\ast &  for $|i-j|=$even \cr
                  g_{j i}^\ast             &       for $|i-j|=$odd \cr
}\eqno(3.16)$$
where   $f_{i j}$ and $g_{i j}$ denote the spin up and down Green's function
in the
configuration space.\par
\smallskip
%%%%%%%%%%%sec34%%%%%%%%%%%%%%%%%%%%%%%%%%%%%%%%%%
\leftline{\it 3-4 general case}\par
We now turn to general case where the lattice size $N$ is $N_1\times N_2$
($N_1$
and $N_2$ stand for the number of lattice sites in the first and second
direction)
and the fermion number of each spin $M=N/2$ ($N=$even).
$M$ fermion state is
$$\eqalign{
c_{m_1,\s}^\dagger c_{m_2 \s}^\dagger \cdot \cdot \cdot
 c_{m_{M} \s}^\dagger\vc
=\sum_{j_1 j_2 ...j_M} \ph_{m_1 j_1}\cdot \cdot \cdot \ph_{m_M j_M}
c_{j_1 \s}^\dagger c_{j_2 \s}^\dagger \cdot \cdot \cdot c_{j_M \s}^\dagger
\vc.
}\eqno(3.17)$$
It changes to
$$\eqalign{
\sum_{j_1 j_2 ...j_M} \ph_{m_1 j_1}\cdot \cdot \cdot \ph_{m_M j_M}
\e_{j_1}\e_{j_2}\cdot \cdot \cdot \e_{j_M}d_{j_1 -\s}\cdot \cdot \cdot d_{j_M
-\s}
\cr
\times {1 \over N!}\ve_{i_1 i_2 ...i_N}d_{i_1 -\s}^\dagger \cdot \cdot \cdot
 d_{i_N -\s}^\dagger
d_{i_1 \s}^\dagger \cdot \cdot \cdot d_{i_N \s}^\dagger \tvc
}\eqno(3.18)$$
Concerning the spin $-\s$ quantities, we have
$$\eqalign{
{{}_NC_M M! \over N!}\sum_{i_1...i_N}\ph_{m_1 i_1}\cdot \cdot \cdot \ph_{m_M
i_M}
\e_{i_1}\e_{i_2}\cdot \cdot \cdot \e_{i_M}
\cr
\times \ve_{i_1 i_2 ...i_M i_{M+1}...i_N}
d_{i_{M+1}}^\dagger \cdot \cdot \cdot d_{i_N }^\dagger\tvc
 \cr
={{}_NC_M M! \over N!}\D \sum_{\{j\} m_{M+1}...m_N}
\tilde\ph_{m_{M+1} j_{M+1}}^\ast\cdot \cdot \cdot \tilde\ph_{m_N j_N}^\ast
\cr
\times \ve_{m_1 ... m_M m_{M+1}...m_N}
d_{j_{M+1}}^\dagger \cdot \cdot \cdot d_{j_N }^\dagger\tvc
}\eqno(3.19)$$
where $\D=$ det$\tilde \ph$, $\tilde \ph_{m j}=\ph_{m j} \e_j$, and ${m_1 ...
m_M}$ are fixed. For dual choice, we have
$$\tilde\ph_{m_{M+1} j_{M+1}}^\ast\cdot \cdot \cdot \tilde\ph_{m_N j_N}^\ast
=\ph_{m_1 j_{M+1}}^\ast\cdot \cdot \cdot \ph_{m_M j_N}^\ast\eqno(3.20)$$
So (3.20) tells us that the %ch
transformed wave functions can be given
by the original wave functions.   Only the difference from the
original wave function
is that they appear in complex conjugate form.  We assume $\vert \D \vert =1$
because of the unitarity of the wave function.
Along with the similar argument given in \S 3-3, we reach
$$
f_{i j}=\cases{
             \d_{i j}-g_{j i}^\ast  &  even spacing \cr
           g_{j i}^\ast                   & odd spacing \cr} \eqno(3.21)
$$
\smallskip
%%%%%%%%%%sec35%%%%%%%%%%%%%%%%%%%%%%%%%%%%%%%%%%%%%
\leftline{\it 3-5 $\chi$SO vs. sum rule}\par
  In order to calculate the chiral spin order parameter ($\chi$SO)\findref
\WWZ,
 six point function is necessary. Six point function with the same spin
operators is
 decomposed into triple products of two point function (Green's function);
$$\eqalign{
\langle c_1^\dagger c_1 c_2^\dagger c_2 c_3^\dagger c_3 \rangle
=&G_{11}G_{22}G_{33}-G_{12}G_{21}G_{33}+G_{12}G_{31}G_{23} \cr
     &- G_{11}G_{23}G_{32}+G_{13}G_{21}G_{32}-G_{13}G_{22}G_{31}  \cr
}\eqno(3.22)
$$
The $\chi$SO is given by
$$E_{123}\equiv \langle  {\vec \s_1} \cdot({\vec \s_2} \times {\vec \s_3)}
\rangle
=-2 i (Pl_{123}-Pl_{132}) \eqno(3.23)$$
where ${\vec \s_i}$ is Pauli matrices sitting at site $i$, and
$$Pl_{123}=\langle \chi_{12}\chi_{23}\chi_{31} \rangle
=\langle c_{1 \s}^\dagger c_{2 \s} c_{2 \s'}^\dagger c_{3 \s'}
 c_{3 \s"}^\dagger c_{1 \s"} \rangle \eqno(3.24)
$$
As is shown in the appendix, $(Pl_{123}-Pl_{132})$ is purely real provided
that
the sum rule (3.21) holds. So a consequence of the sum rule is that the real
part
of the $\chi$SO
$$ {\rm Re} E_{123}=2 {\rm Im} (Pl_{123}-Pl_{132}) \eqno(3.25)$$
becomes zero. \par
  This holds only for the half-filling case and does not tell anything about
doped cases.
In the following section we will explicitly  see the above properties  by
making numerical simulations.

\bigskip
%
%\vfill\eject
%%%%%%%%%%%%%%%sec4%%%%%%%%%%%%%%%%%%%%%%%%%%%%%%%%%%%%%%%
\leftline{\bf \S 4. Monte Carlo simulations}\par
In the present paper we take $4\times 4 $ spatial lattice, and the Trotter
size $L$
 is taken
 to be 100 through  200, so that $\D \t$ is sufficiently small ( $\D \t \leq
0.1$).
 The hopping parameter $t$ is kept to be 1, and $U$ is  taken to be
 4.0.  The wave function $\lbrace \phi_{m i}
\rbrace$  is  on the free particle basis in the  complex form $\exp(i { 2\pi
\over
 4} {\bf k \cdot n})$.  We choose particle states with   momentum ${\bf k }$
 so as to fill the energy states   in order from the lowest  to higher  ones.
We make a dual choice of section 3.  For half-filling, the lower-half energy
states
are occupied by $\lbrace \phi_{m i} \rbrace$ and the dual counterparts
$\lbrace \psi_{m i} \rbrace$  correspond  to the upper-half. Setting
$\lbrace \phi_{m i} \rbrace=\exp [i 2\pi {\bf k \cdot n}/4]$, our choice is
${\bf k}=(k_1,k_2)=(0,0)$  for $\phi_m=\phi_1$, $(1,0)$ for $\phi_2$,
$(-1,0)$ for $\phi_3$,    $(0,1)$ for $\phi_4$,     $(0,-1)$ for $\phi_5$,
$(1,1)$ for $\phi_6$ and $(1,-1)$ for $\phi_7$. As for  $\phi_8$, we choose
a linear combination of the two states  $(2,0)$ and $(0,2)$.
Dual counterparts are chosen in such a way that
$ (k_1,k_2)=(2,2)$ for $\psi_1$,
$(-1,2)$ for $\psi_2$, $(1,2)$ for $\psi_3$, $(2,-1)$ for $\psi_4$,
 $(2,1)$ for $\psi_5$, $(-1,-1)$ for $\psi_6$,
$(-1,1)$ for $\psi_7$ and $\psi_8=\phi_8$.
In Fig.1, we show them
in the momentum space.  For spin up and down sectors we take the same wave
 functions $\lbrace \phi_{m i} \rbrace$, and $m$ runs from 1 to $\M$.  We
performed
from 3000 to 18000 sweeps depending on $\M$ (the number of fermions with each
spin
 $\s$ (2.11)) and $\b$. The statistical errors are
estimated by use of the block method with each block size being 500 to 1000
after 1000 warming-up sweeps.
\smallskip
%%%%%%%%%%%%%sec41%%%%%%%%%%%%%%%%%%%%%%%%%%%%%%%%%%%%%%%5555
\leftline{\it 4.1 half filled case ($\M=8$)}\par
For the half filled case, no sign problem is involved,  since the average of
phase
is purely real and is unity for any configurations occurred in the algorithm.
We define symmetrized two point functions
$$
f_{\lbrace  i j \rbrace} \equiv {1 \over 2}(f_{i j} + f_{j i})  \qquad
g_{\lbrace  i j \rbrace} \equiv {1 \over 2}(g_{i j} + g_{j i})
\eqno(4.1)
$$
Spin average of the symmetrized functions satisfy
$$\eqalign{
&{\rm Re} (f_{\lbrace  i j \rbrace}+g_{\lbrace  i j \rbrace})=0  \qquad  {\rm
for \ }  |i-j|={\rm
even} \neq 0\cr
&{\rm Re} (f_{\lbrace  i i \rbrace}+g_{\lbrace  i i \rbrace})=1.0 \qquad
\cr
&{\rm Im} (f_{\lbrace  i j \rbrace}+g_{\lbrace  i j \rbrace})=0   \qquad {\rm
for \ }  |i-j|= {\rm
odd} \cr
}\eqno(4.2)
$$
In Fig.2a, we show results of the Green's function, and compare with those of
the Lanczos's method\findrefs \Paroa \Parob \endrefs.  The average values of
the real
part for even spacings ($\vert i-j \vert \neq 0$)
are exactly zero in
agreement with the argument in section 3.
Fig.2b is spin-spin correlations. It shows clear antiferromagnetic
correlations
 and agrees very well with the Lanczos's one. The  numerical figures for the
Green's functions and
 the spin-spin correlations  are listed in Table \Rn(1) .\par
The real part of $\chi$SO turns out exactly vanishing as discussed in the
previous section. We  looked also at the  imaginary
part, and it  is consistent with zero within error.
\smallskip
%%%%%%%%%sec42%%%%%%%%%%%%%%%%%%%%%%%%%%%%%%%%%%
\leftline{\it 4.2   sign problem }\par
For simulations of doped cases one is confronted with the sign problem.
 In our case, for the complex trial wave function,  $W$ is generally complex,
although $\rho(\b;\Phi)$ (2.3) is real.
As stated in section 2, one generates a sequence of configurations with a
probability\findref \Loh, \par
$$P_+({s})={\vert W({s})\vert  \over \rho_+} \eqno(4.3)$$
where
$$\rho_+=\sum_{{s}}\vert W({s})\vert\eqno(4.4)$$
 The average of the phase of $W$, which appears in (2.22), is then given by
$$   <e^{i \theta_w}>_+= \r/\r_+.\eqno(4.5)$$
For   large $\b$,    $\r$ is dominated by the ground state $\vert  \psi_0
\rangle$
with energy $E_0$
$$ \r \simeq \exp (-\b E_0) \vert <\psi_0|\Phi>\vert ^2\eqno(4.6),$$
while $\rho_+$ looks like
$$ \r_+ \simeq \exp (-\b E_+)\vert <\psi_{+0}|\Phi>\vert ^2,\eqno(4.7)$$
where $E_+$ and  $\psi_{+0}$ are analogue of $E_0$ and $\psi_0$
 for $\r_+$, respectively.
 Therefore the average of the phase is real,
and   behaves in a similar manner
to that of the sign in the real trial wave function case;
 $$  \langle e^{i \theta_w} \rangle_+= \r/\r_+ \simeq \exp(-\b \D E)
\eqno(4.8)$$
 where $\D E \equiv E_0-E_+$.
 For $\D E=0$,  the average of sign converges to
 some finite value.  In this case, one has no difficulty with the phase, and
one may
 neglect the phase for evaluating the average of
physical quantities.
   For $\D E \neq 0$, on the other hand, we encounter
 so called the sign problem, that is, $\langle e^{i \theta_w} \rangle$
 becomes vanishing in our case.
It apparently gives meaningless result for the average of any physical
  operator (2.22) and   anticipates, at the same time,   large errors.
If, however, the numerator in (2.22) behaves in a exponential manner
similarly
  to the denominator,
  it is still expected that the average may converge to some finite value,
  and its  errors could be tamable.
This depends upon  the fillingness and operators to be calculated.
In the following subsections we show the results concerned with this issue
for $\M$=7 and $\M$=5 (closed shell).
\smallskip
%%%%%%%%sec43%%%%%%%%%%%%%%%%%%%%%%%%%%%%%%%%%%%
\leftline{\it 4.3 Hole doped cases $\M=7$}\par
The average of the real part of  the phase is shown in Fig.3 and Table \Rn(2)
{}.
 In the region we have calculated ($\beta= 2.0$ to $5.0$),
 an exponential behavior
 is clearly observed.  The $\D E$ read from the slope is 1.01(5).
This value is somewhat larger than the one obtained by Loh et al\findref
\Loh.
 The average  $<O>_+ $ with the measure (2.21)  for   Green's functions
($O=G_{i j}$)
and for spin correlations
 are determined in a  high precision, while the real part of the
$O=\chi$SO
 parameter is somewhat noisy and looks unstable as $\b$ varies.
  This difference of the fluctuations in $<O>_+ $  results in   the
difference of weighted
averages (2.22)  among the Green's function, spin-spin correlations and the
$\chi$SO.\par
According to the formula (2.22), the average of  observables is proportional
to
 the product of observables and  the phase  in the  numerator.
We have found  that due to the steadiness of behaviors of the Green's
function $G_{ij}$  and
spin-spin correlations with respect to
a variation of $\b$, their product shows an exponential fall-off with almost
the same
slope as the denominator $\langle e^{i \theta_w} \rangle_+$ .
 On the other hand,  the product of $\chi$SO and phase does not
show such behavior. Fig.4a and 4b show   the real and the imaginary  parts of
the  $\chi$SO as
a function of $\b$.
We see that for small $\b$ ($\b=2.0$ and 3.0, for example), the real part has
 some
 positive value\findref \Paroa within error,
 but becomes unfortunately too noisy to extrapolate to the larger $\b$
region (see Table \Rn(2) ).
We should note that the choice of the wave function in Fig.1 explicitly
breaks the parity.
This  could  be   the reason why the $\chi$SO may be non-vanishing for small
$\b$.  It is then
important to study how the operator behaves in the large $\b$ limit.
Since the spontaneous symmetry breaking never happens in a finite volume
in the strict sense \findref \HatS ,    the $\chi$SO goes to zero as $\b$
increases.
As in the case of an introduction of an  external source to break explicitly
a symmetry, we must study  finite size effects how the operator becomes
 vanishing.  This is under investigation.\par
    We have got better results as to the Green's function and the spin-spin
correlations.
 Fig.5 compares them with
the Lanczos method.
According to  Parola et al.\findrefs \Paroa \Parob \endrefs ,  the ground
states are found to be threefold
degenerate; one is the d-wave state of momentum (0,0)  and the others  are  a
pair
 of  (0,$\pi$) and ($\pi$,0). Both the states show slightly different
behavior
for the density distribution. The Monte Carlo results are  in good agreement
with
that for (0,0).  Spin-spin correlations are shown in Fig.6.  The data used
for drawing the figures
are listed in Table \Rn(1) .

\smallskip
%%%%%%%%%sec44%%%%%%%%%%%%%%%%%%%%%%%%%%%%%%%%%%%%%%%%%
\leftline{\it 4.4 Hole doped cases $\M=5$}\par
  The case  $\M=5$ makes  a ``closed shell'' in the single free particle
distribution
in the momentum space.   The sign problem is very much mild compared to the
$\M =7$ case.
  The average of the real part of the phase is shown in Fig.7 and Table
\Rn(2) .   One sees an
 exponential fall-off behavior for a wide range of $\b$ with  a  decay rate
being
quite small compared to
the $\M=7$ case; $\D E=0.0018(3) $.    We see then that $\D E$ depends much
on the fillingness.
The unweighted Green's function as a
function of $\b$ are very stable and gets very small errors.
  These facts  lead us to expect that the weighted average is approximately
 given by the
unweighted one. In other words, a  correction to the unweighted average is
 quite small.
In  Fig.8,  we compare them with the ones obtained  by  the Lanczos method.
We see a very good agreement.   The spin-spin correlations ( Fig.9.) are also
in very good agreement with the exact numerical calculations.\findref \Parob
The numerical figures are listed in Table \Rn(1) for the Green's function and
the spin-spin
 correlations.
 As for the behavior of $\chi$,  both the real and imaginary parts are
consistent
 with   zero within   error.
\bigskip
%%%%%%%sec5%%%%%%%%%%%%%%%%%%%%%
\leftline{\bf \S 5. Summary and Discussion}\par
We performed Monte Carlo calculations of the two dimensional Hubbard model
by use of complex wave functions as a trial state.
As to the single  particle density
operators, or the Green's functions and spin-spin correlations, we have seen
good agreements with the Lanczos' results for half-filled and doped cases.
Encouraged by this fact, we tried to measure the $\chi$SO.
For half filled case, without numerical calculations, we have seen that it is
vanishing
 based upon the particle-hole transformation.
Upon doping, the %ch
phase problem is a main obstacle.   For $\M=7$, the problem is rather
 serious, and we are not able to extrapolate to large enough values of $\b$,
 though the real part of the $\chi$SO looks nonvanishing at small $\b$.
In order to clarify it, we need to study  finite size effect to see how
this operator reaches zero, since no spontaneous
symmetry breaking occurs in finite volume in the strict sense.
  This issue is
under investigation.  For $\M=5$, %ch
the phase problem is very mild.  In this case,
 the real part of the $\chi$SO is consistent with zero independent of values
of $\b$.
As to  the average of plaquettes, which is a  product
of four links $\langle \chi_{1 2}\chi_{2 3} \chi_{3 4} \chi_{4 1} \rangle$,
we will
report in the forthcoming paper.

\par
\bigskip
%%%%%%%%%%%%%%%%%%ack%%%%%
\leftline {\bf{ACKNOWLEDGMENTS}}
  We are grateful to Dr. N. Hatano for discussions concerning
the measurement of
the $\chi$SO.
 %ch
 We appreciate the collaboration of K. Yamaji in the early stage of this
work. We are very much indebted to the Illinois University, where
the preliminary calculations were performed,  and  to the RCNP,
where the main calculations have been done.
\bigskip
%%%%%%%%%%%%%%%%ref%%%%%%%%%%%%%
\outrefs
\vfill\eject
%\smallskip
%%%%%%%%%%%%tabc%%%%%%%%%%%%%%%%%%%%%%%
\leftline\bf{TABLE  CAPTIONS}
\item{Table  \Rn(1) }The average values and the statistical errors of the
single particle
 density operators (spin averaged symmetrical Green's function) $G_i$ and the
spin-spin correlations
$S_i$ for various fillings.  Suffix $i$ stands for the locations of the sites
along the path
as depicted in the inset of Fig.2a.  The values  for $\M=8, \M=7$ and $\M=5$
are shown at  $\b=5.0,
 \b=4.0$ and $\b=20.0$, respectively. Note that $G_0=1.0$ for $\M=8$.
%%%
\item{Table \Rn(2) }The real part of the average of the phase vs. $\b$ for
$\M=7$ and $\M=5$.
The real and imaginary parts of the $\chi$SO is also listed  for $\b=7$.
Meas stands for
the number of measurements (1$k=$ 1000).
\bigskip
%%%%%%%%%%%%fig%%%%%%%%%%%%%%%%%%%%%%%%%%%%
\leftline\bf{FIGURE CAPTIONS}
%\smallskip
%\parindent -10pt{
\item{Fig. 1 }  Illustration of the ``Dual choice'' of wave functions in the
momentum
space.  $\lbrace \phi_{m} \rbrace$ and $\lbrace \psi_ m \rbrace$  are shown
(see  text).
\item{Fig. 2a} Expectation value of single particle density operator, or spin
averaged
symmetrical Green's function $G_i$ for half-filling and $\b=5.0$.
 The horizontal axis stands for the locations $i$
 along the path  shown in the inset.  The first and  third   sites are odd,
and the rests are
even spacings.
Circle is the results of the Monte Carlo calculations, while cross is  the
Lanczos'
 results\findref \Parob.
Lines are drawn for the guide of eyes.
\item{Fig. 2b} Spin-spin correlations $S_i$ indicate clear antiferromagnetic
correlations
for half filling and $\b=5.0$. The horizontal axis stands for the same
location $i$ as in
Fig.2a. Monte Carlo calculations ( circle) are in good agreement with those
of
Lanczos' method (cross)\findref \Parob.  The errors lie within the symbols.
\item{Fig. 3 } Average of the real part of the phase of the Monte Carlo
weights
 vs. $\b$ for $\M =7$.
\item{Fig. 4a}Real part of the $\chi$SO vs. $\b$ for $\M=7$.
\item{Fig. 4b}Imaginary part of the $\chi$SO vs. $\b$ for $\M=7$.
\item{Fig. 5 }Single particle density operator $G_i$ for $\M=7$ and $\b$=4.0.
We compare the Monte Carlo results (circle) with the Lanczos' ones\findref
\Parob.
According to Parola et. al.\findref \Parob, the degenerate ground states (see
text) show
different behavior.   Monte Carlo results agree well with the behavior
of the  (0,0)  state (cross). The symbol (square) stands for the behavior of
(0,$\pi$)
 and ($\pi$,0) state.
\item{Fig. 6 }Spin-spin correlations $S_i$ for $\M=7$ and $\b$=4.0.  Monte
Carlo results (circle)
  agree with
the Lanczos' results\findref \Parob ( cross and square).
\item{Fig. 7 }The real part of the phase of the Monte Carlo weights vs. $\b$
for $\M =5$.
An exponential fall-off behavior is seen in the region where $\b \geq 10.0$.
\item{Fig. 8 }Single particle density operator $G_i$ for $\M=5$ and
$\b$=20.0.  The Monte Carlo
results (circle) are in very good agreement with those of Lanczos'
method\findref \Parob (cross).
The errors lie  within the symbols.
\item{Fig. 9 }Spin-spin correlations $S_i$ (circle) for $\M=5$ and $\b$=20.0.
 Comparison with the
Lanczos' results\findref \Parob
 (cross). The errors lie  within the symbols.
%}
%\TeX\footnote{*}{ \TeX\ is a trade mark }
\vfill\eject
%%%%%%%%%%%%%%%%%app%%%%%%%%%%%%%%%%%%%%%%%%
\leftline\bf{APPENDIX}\par
In this appendix, we show that $(Pl_{123}-Pl_{132})$ appearing in (3.23) is
pure real.
$\chi$SO is expressed by two through six point functions.
Four point functions are
$$\eqalign{
 \langle \bar 1 1 \bar 2 2 \rangle_\s
 &\equiv \langle c_{1 \s}^\dagger c_{1 \s} c_{2 \s}^\dagger c_{2 \s} \rangle
             = G_{1 1}^{(\s)}G_{2 2}^{(\s)}-G_{1 2}^{(\s)}G_{2 1}^{(\s)} \cr
\langle \bar 1 2 \bar 2 3 \rangle_\s
&\equiv  \langle c_{1 \s}^\dagger c_{2 \s} c_{2 \s}^\dagger c_{3 \s} \rangle
=G_{3 1}^{(\s)}-G_{3 1}^{(\s)}G_{2 2}^{(\s)}+G_{3 2}^{(\s)}G_{2 1}^{(\s)}
}\eqno(A.1)$$
where $G_{i j}^{(\s)}=\langle  c_{j \s}^\dagger c_{i \s}\rangle$ for spin
$\s$.
$(Pl_{123}-Pl_{132})$  in (3.23)
$$\eqalign{
Pl_{123}-Pl_{132}
=\sum_{\s, \s', \s''} \lbrace
\langle c_{1 \s}^\dagger c_{2 \s} c_{2 \s'}^\dagger c_{3 \s'}c_{3
\s''}^\dagger c_{1 \s''} \rangle-
\langle c_{1 \s}^\dagger c_{3 \s} c_{3 \s'}^\dagger c_{2 \s'}c_{2
\s''}^\dagger c_{1 \s''} \rangle
\rbrace
}\eqno(A.2)$$
is written as
$$\eqalign{
Pl_{123}-Pl_{132}
&=\langle \bar 1 1 \bar 2 2 \bar 3 3\rangle_\uparrow
  -\langle \bar 1 1 \bar 3 3 \bar 2 2\rangle_\uparrow
  +\langle \bar 1 2 \bar 2 3\rangle_\uparrow \langle  \bar 3
1\rangle_\downarrow
  -\langle \bar 1 3 \bar 3 2\rangle_\uparrow \langle \bar 2 1
\rangle_\downarrow\cr
&+\langle \bar 1 2 \bar 3 1\rangle_\uparrow \langle  \bar 2
3\rangle_\downarrow
 -\langle \bar 1 3 \bar 2 1\rangle_\uparrow \langle \bar 3 2
\rangle_\downarrow
 +\langle \bar 2 3 \bar 3 1\rangle_\uparrow \langle  \bar 1
2\rangle_\downarrow
 -\langle \bar 1 3 \rangle_\uparrow \langle \bar 3 2 \bar 2 1
\rangle_\downarrow\cr
&+\langle \bar 1 2 \rangle_\uparrow \langle \bar 2 3 \bar 3 1
\rangle_\downarrow
 -\langle \bar 1 3 \rangle_\uparrow \langle \bar 3 2 \bar 2 1
\rangle_\downarrow
 +\langle \bar 2 3 \rangle_\uparrow \langle \bar 1 2 \bar 3 1
\rangle_\downarrow
 -\langle \bar 3 2 \rangle_\uparrow \langle \bar 1 3 \bar 2 1
\rangle_\downarrow\cr
&+\langle \bar 3 1 \rangle_\uparrow \langle \bar 1 2 \bar 2 3
\rangle_\downarrow
 -\langle \bar 2 1 \rangle_\uparrow \langle \bar 1 3 \bar 3 2
\rangle_\downarrow
+\langle \bar 1 1 \bar 2 2 \bar 3 3\rangle_\downarrow
-\langle \bar 1 1 \bar 3 3 \bar 2 2\rangle_\downarrow\cr
&=-f_{2 2}f_{3 1}g_{1 3} + f_{2 2}f_{1 3}g_{3 1}+g_{2 2}f_{3 1}g_{1 3}-g_{2
2}f_{1 3}g_{3 1}\cr
&+(\rm{ terms\; without\;} f_{2 2}  \; \rm{or} \; g_{2 2})
}\eqno(A.3)$$
where we  used the notation
$ f_{i j}\equiv G_{i j}^{(\uparrow)}$ and  $g_{i j}\equiv G_{i
j}^{(\downarrow)} $.
The sum rule (3.21) and (A.3) lead to
$$\eqalign{
Pl_{123}-Pl_{132}\cr
&=(g_{1 3}^\ast g_{1 3} - g_{3 1}^\ast g_{3 1})
- (g_{2 2}^\ast + g_{2 2})(g_{1 3}^\ast g_{1 3} - g_{3 1}^\ast g_{3 1})\cr
&+ (\rm { terms\;  without\;} f_{2 2} \; \rm{or} \; g_{2 2}) \cr
&=\rm{pure \; real} + (\rm{terms \; without}\; f_{2 2} \;\rm{or} \; g_{2 2})
}\eqno(A.4)$$
The terms without $f_{2 2}$ or $g_{2 2}$ in (A.4) can also be rewritten by
the sum rule
(3.21) in the form as sum of terms $\alpha+\alpha^\ast$ and $\b \b^\ast$,
where $\alpha$
and $\b$ are complex numbers.
\bye